\documentclass{article}

\usepackage{microtype}
\usepackage{graphicx}
\usepackage{subcaption}
\usepackage{booktabs} 

\usepackage{hyperref}

\usepackage{amsmath}
\usepackage{amssymb}
\usepackage{mathtools}
\usepackage{amsthm}
\usepackage{multirow}
\usepackage{multicol}
\usepackage{enumitem}
\usepackage{makecell}

\usepackage[preprint]{icml2026}

\usepackage{amsmath}
\usepackage{amssymb}
\usepackage{mathtools}
\usepackage{amsthm}

\usepackage[capitalize,noabbrev]{cleveref}

\theoremstyle{plain}

\theoremstyle{definition}

\theoremstyle{remark}

\usepackage{xcolor,comment}

\newcommand{\hide}[1]{}

\usepackage[textsize=tiny]{todonotes}

\icmltitlerunning{One Brain, Omni Modalities: Towards Unified Non-Invasive Brain Decoding with Large Language Models}

\begin{document}

\twocolumn[
\icmltitle{One Brain, Omni Modalities: Towards Unified Non-Invasive Brain Decoding with Large Language Models}





  \icmlsetsymbol{equal}{*}

  \makeatletter
  \setcounter{@affiliationcounter}{3}
  \newcounter{@affilailab}
  \setcounter{@affilailab}{1}
  \newcounter{@affilstu}
  \setcounter{@affilstu}{2}
  \newcounter{@affilthu}
  \setcounter{@affilthu}{3}
  \makeatother

  \begin{icmlauthorlist}
    \icmlauthor{Changli Tang}{thu}
    \icmlauthor{Shurui Li}{ailab,stu}
    \icmlauthor{Junliang Wang}{ailab}
    \icmlauthor{Qinfan Xiao}{ailab,thu}
    \icmlauthor{Zhonghao Zhai}{thu}
    \icmlauthor{Lei Bai}{ailab}
    \icmlauthor{Yu Qiao}{ailab}
    \icmlauthor{Bowen Zhou}{ailab,thu}
    \icmlauthor{Wen Wu}{ailab}
    \icmlauthor{Yuanning Li}{ailab,stu}
    \icmlauthor{Chao Zhang}{ailab,thu}
  \end{icmlauthorlist}

  \icmlaffiliation{thu}{Tsinghua University}
  \icmlaffiliation{ailab}{Shanghai AI Laboratory}
  \icmlaffiliation{stu}{ShanghaiTech University}

  \icmlcorrespondingauthor{Chao Zhang}{cz277@tsinghua.edu.cn}

  \icmlkeywords{Machine Learning, ICML}

  \vskip 0.3in
]



\printAffiliationsAndNotice{}  

\begin{abstract}
Deciphering brain function through non-invasive recordings requires synthesizing complementary high-frequency electromagnetic (EEG/MEG) and low-frequency metabolic (fMRI) signals. However, despite their shared neural origins, extreme discrepancies have traditionally confined these modalities to isolated analysis pipelines, hindering a holistic interpretation of brain activity. To bridge this fragmentation, we introduce \textbf{NOBEL}, a \textbf{n}euro-\textbf{o}mni-modal \textbf{b}rain-\textbf{e}ncoding \textbf{l}arge language model (LLM) that unifies these heterogeneous signals within the LLM's semantic embedding space. Our architecture integrates a unified encoder for EEG and MEG with a novel dual-path strategy for fMRI, aligning non-invasive brain signals and external sensory stimuli into a shared token space, then leverages an LLM as a universal backbone. Extensive evaluations demonstrate that NOBEL serves as a robust generalist across standard single-modal tasks. We also show that the synergistic fusion of electromagnetic and metabolic signals yields higher decoding accuracy than unimodal baselines, validating the complementary nature of multiple neural modalities. Furthermore, NOBEL exhibits strong capabilities in stimulus-aware decoding, effectively interpreting visual semantics from multi-subject fMRI data on the NSD and HAD datasets while uniquely leveraging direct stimulus inputs to verify causal links between sensory signals and neural responses. NOBEL thus takes a step towards unifying non-invasive brain decoding, demonstrating the promising potential of omni-modal brain understanding.
\end{abstract}

\section{Introduction}
The human brain operates as an intricate biological processor, orchestrating complex neural activities that encode our perceptions, thoughts, and physiological states. To non-invasively decipher these mechanisms, researchers primarily rely on two distinct categories of signals: high-frequency electromagnetic fluctuations arising from neuronal firing, captured via electroencephalography (EEG) and magnetoencephalography (MEG), and low-frequency metabolic changes reflecting hemodynamic responses, captured via functional magnetic resonance imaging (fMRI). Driven by the success of representation learning, recent years have witnessed a surge in foundation models tailored for these individual modalities. Significant progress has been made in building specialized models for EEG~\cite{xiao2025brainomni, peh2022transformer, song2021transformer, jing2023development, yang2023self, jiang2024labram, wang2025cbramod, ouahidi2025reve, yi2023learning}, MEG~\cite{xiao2025brainomni, hirano2024deep}, and fMRI~\cite{wang2025towards, scotti2024mindeye, kawahara2017brainnetcnn, li2021braingnn, kan2022brain, kan2022fbnetgen, wei2025fmri, dong2025brain, kim2023swift, ozcelik2023natural, scotti2023reconstructing, chen2025mindgpt, tang2023semantic, horikawa2025mind, fu2025making, takagi2023high}, each demonstrating strong capabilities in decoding specific aspects of brain function.

Despite these advances, the field remains fragmented. While electromagnetic and metabolic signals are merely different manifestations of the same underlying neural processes, they are traditionally treated as isolated domains due to their extreme heterogeneity. EEG and MEG offer millisecond-level temporal resolution but poor spatial localization, whereas fMRI provides rich spatial details but suffers from significant temporal lag. Theoretically, a unified analysis of these complementary modalities could unlock a holistic view of brain dynamics, yet current frameworks lack the capability to bridge this ``frequency gap'' effectively. Furthermore, neural activity does not occur in a vacuum; it is deeply grounded in the external sensory environment. However, most of recent brain foundation models aim to learn task-agnostic neural representations and therefore do not explicitly incorporate task-specific or stimulus-level contextual information. As a result, the rich structure of the evoking stimuli is often underutilized during representation learning. Integrating these external stimuli as contextual input is crucial, as it provides the ground truth necessary to align complex brain signals with high-level semantic concepts, thereby raising the ceiling of decoding performance.

To address these challenges, we introduce NOBEL, a neuro-omni-modal brain-encoding LLM designed to bridge the chasm between high-frequency electromagnetic activities, low-frequency metabolic states, and external sensory stimuli. We conceptualize brain signal decoding as a multi-modal alignment problem, leveraging the strong LLMs to integrate heterogeneous data sources. Specifically, NOBEL employs a modular encoding strategy to project diverse signals into a shared semantic space compatible with the LLM. For high-frequency data,  BrainOmni \cite{xiao2025brainomni} is used to extract temporal features from EEG and MEG. For fMRI signals, we propose a novel dual-path encoding mechanism, employing NeuroSTORM \cite{wang2025towards} to capture global, static subject-dependent meta-information and the MindEye2 structure \cite{scotti2024mindeye} to extract dynamic, stimulus-evoked perceptual features. By simultaneously ingesting these brain signals alongside the corresponding visual or auditory stimuli, NOBEL translates multi-modal inputs into unified tokens, allowing users to control decoding tasks and generate insights via natural language prompts.

Our contributions are summarized as follows:
\begin{itemize}[itemsep=0pt, leftmargin=*]
    \item \textbf{First Omni-Modal Brain Foundation Model:} We present NOBEL, a unified framework that supports EEG, MEG, and fMRI processing. By orchestrating these heterogeneous signals within a single LLM, NOBEL achieves competitive performance across a wide range of benchmarks compared to modality-specific specialists.

    \item \textbf{Novel Dual-Path fMRI Encoding:} We propose a hybrid encoding architecture specifically optimized for metabolic signals to disentangle the extraction of static physiological meta-information from dynamic, stimuli-responsive features, which effectively capturing the rich spatiotemporal information embedded in low-frequency fMRI data.
    \item \textbf{Stimuli-Aware Task Generalization:} We demonstrate that NOBEL not only effectively performs stimuli decoding tasks but is also able to leverage external audio-visual stimuli as contextual anchors to tackle tasks beyond the reach of previous models, such as verifying whether a subject is perceiving a specific target image or video.
\end{itemize}

The rest of the paper is structured in the following manner. Section~\ref{sec:related_work} reviews existing research and foundation models for EEG/MEG and fMRI signals. Section~\ref{sec:method} details the methodology of NOBEL, describing the unified model architecture and the multi-stage training pipeline. Section~\ref{sec:setup} provides the experimental setup, including data specifications and implementation details. Section~\ref{sec:expres} presents comprehensive experimental results. We conclude in Section~\ref{sec:conclude}.

\section{Related Work}
\label{sec:related_work}
\subsection{Foundation Models for EEG and MEG Signals}
\label{sec:related_eeg_meg}

The analysis of non-invasive high-frequency electromagnetic brain signals, specifically EEG and MEG, has historically been challenged by low signal-to-noise ratios (SNRs) and the complex spatial-temporal dynamics of neuronal firing. 
Early deep learning approaches predominantly focused on designing specialized architectures to tackle specific clinical tasks. For instance, CNN-Transformer~\cite{peh2022transformer} integrated CNNs for local feature extraction with Transformers for global context to automate artifact detection, while ST-Transformer~\cite{song2021transformer} utilized a similar hybrid structure to decode spatial-temporal features from EEG. In clinical settings, SPaRCNet~\cite{jing2023development} demonstrated that dense convolutional networks could achieve expert-level performance in classifying seizures and rhythmic periodic patterns. Similarly, in the MEG domain, FAMED~\cite{hirano2024deep} employed deep learning for the automatic detection and dipole estimation of epileptic discharges, proving the efficacy of deep networks in magnetic signal analysis.

Despite the success of these supervised methods, their reliance on extensive annotated data limited their scalability. This led to a shift towards self-supervised learning (SSL), where models learn robust representations from unlabeled data. MMM \cite{yi2023learning} maps any EEG channel distribution into a unified topology and conducts multi-stage pretraining. ContraWR~\cite{yang2023self} applies contrastive learning to EEG, enabling automated sleep staging with reduced reliance on labeled experts. LaBraM~\cite{jiang2024labram} further scaled pre-training through vector-quantized neural spectrum prediction. More recently, CBraMod~\cite{wang2025cbramod} introduced a criss-cross attention mechanism to better capture the orthogonal relationships between spatial and temporal dimensions in EEG data. REVE~\cite{ouahidi2025reve} utilized a 4D positional encoding scheme to process signals of any length and electrode arrangement.

Notably, neuronal activity underpins human brain function by generating electrical currents in the cortex, which in turn produce secondary electrical and magnetic fields measurable by EEG and MEG. Although these signals stem from the same underlying neural sources, previous foundation models have largely ignored this biophysical unity and treated electric and magnetic modalities as separate domains. The recently proposed BrainOmni~\cite{xiao2025brainomni} bridges this gap by mapping diverse sensor configurations into a shared latent space and establishes a unified EMEG encoder.

\subsection{Foundation Models for fMRI Signals}
\label{sec:related_fmri}

While EEG and MEG capture rapid neural dynamics, fMRI provides spatially resolved metabolic insights. The development of deep learning models for fMRI has largely been dictated by data representation strategies. Traditionally, to mitigate the curse of dimensionality, fMRI data is reduced to region-of-interest (ROI)-level time-series extracted from pre-defined brain atlases. From these time-series, various low-dimensional matrix representations like functional connectivity or functional gradients can be derived, which summarize inter-regional relationships. Early supervised models treated these matrix representations as grid-like structures or graphs; for instance, BrainNetCNN~\cite{kawahara2017brainnetcnn} pioneered the use of specialized convolutional filters to capture topological features, while subsequent works like BrainGNN~\cite{li2021braingnn} and FBNETGEN~\cite{kan2022fbnetgen} adopted graph neural networks to improve interpretability and task-aware network generation. With the advent of Transformers, architectures such as the Brain Network Transformer~\cite{kan2022brain} were proposed to learn pair-wise connection strengths among ROI by utilizing connection profiles as node features and incorporating an orthonormal clustering readout. Recent efforts have attempted to scale this paradigm to foundation model settings, with Brain Harmony~\cite{dong2025brain} aligning multi-site data via unified ROI embeddings and fMRI-LM~\cite{wei2025fmri} exploring the intersection of ROI sequences and language models. However, these ROI-based methods suffer from a fundamental bottleneck: the reliance on fixed atlases introduces rigid inductive biases and inconsistent preprocessing pipelines, limiting their ability to generalize across datasets processed with different parcellation schemes.


In contrast, directly modeling raw 4D voxel data offers a promising alternative that avoids manual parcellation and preserves fine-grained neural information. 
This volume-based paradigm treats fMRI as dense spatiotemporal videos rather than simplified graphs. SWiFT~\cite{kim2023swift} successfully applied Swin Transformers to 4D fMRI clips, showing the feasibility of learning hierarchical spatiotemporal representations without manual feature engineering. Pushing this direction further, NeuroSTORM~\cite{wang2025towards}  extracts robust, fine-grained voxel features that generalize well to downstream tasks such as phenotype prediction and disease diagnosis, by pre-training on massive multi-subject datasets using masked autoencoding.

Distinct from methods that primarily focus on analyzing physiological states, a parallel stream of research aims to explicitly decode perceptual content, such as decoding visual or auditory stimuli from brain activity. 
For decoding continuous stimuli (\textit{e.g.} long natural speech or videos), models \cite{tang2023semantic,fu2025making} typically operate directly on the fMRI time series to translate fMRI signals into perceived continuous stimuli.
In contrast, for event-related stimuli, event-related neural representations are commonly used, such as general linear model (GLM)-derived beta weights \cite{takagi2023high} or trial-averaged activations \cite{chen2025mindgpt,horikawa2025mind}, focusing on the semantic alignment between brain activity and external stimuli. Generative approaches like BrainDiffuser~\cite{ozcelik2023natural} and MindEye~\cite{scotti2023reconstructing} are also based on beta weights and have achieved superior results in reconstructing perceived images by aligning fMRI embeddings with the latent spaces of diffusion models such as CLIP \cite{radford2021learning}. Similarly, NeuroBind~\cite{yang2024neurobind} leverages CLIP embeddings as a universal anchor to unify diverse neural modalities into a shared semantic space. The recently proposed MindEye2~\cite{scotti2024mindeye} further enhances the neural-visual alignment, enabling high-fidelity reconstruction with less training data. 


\section{Methodology}
\label{sec:method}
\subsection{Model Architecture}
\label{sec:architecture}

\begin{figure*}[ht]
\begin{center}
\centerline{\includegraphics[width=\linewidth]{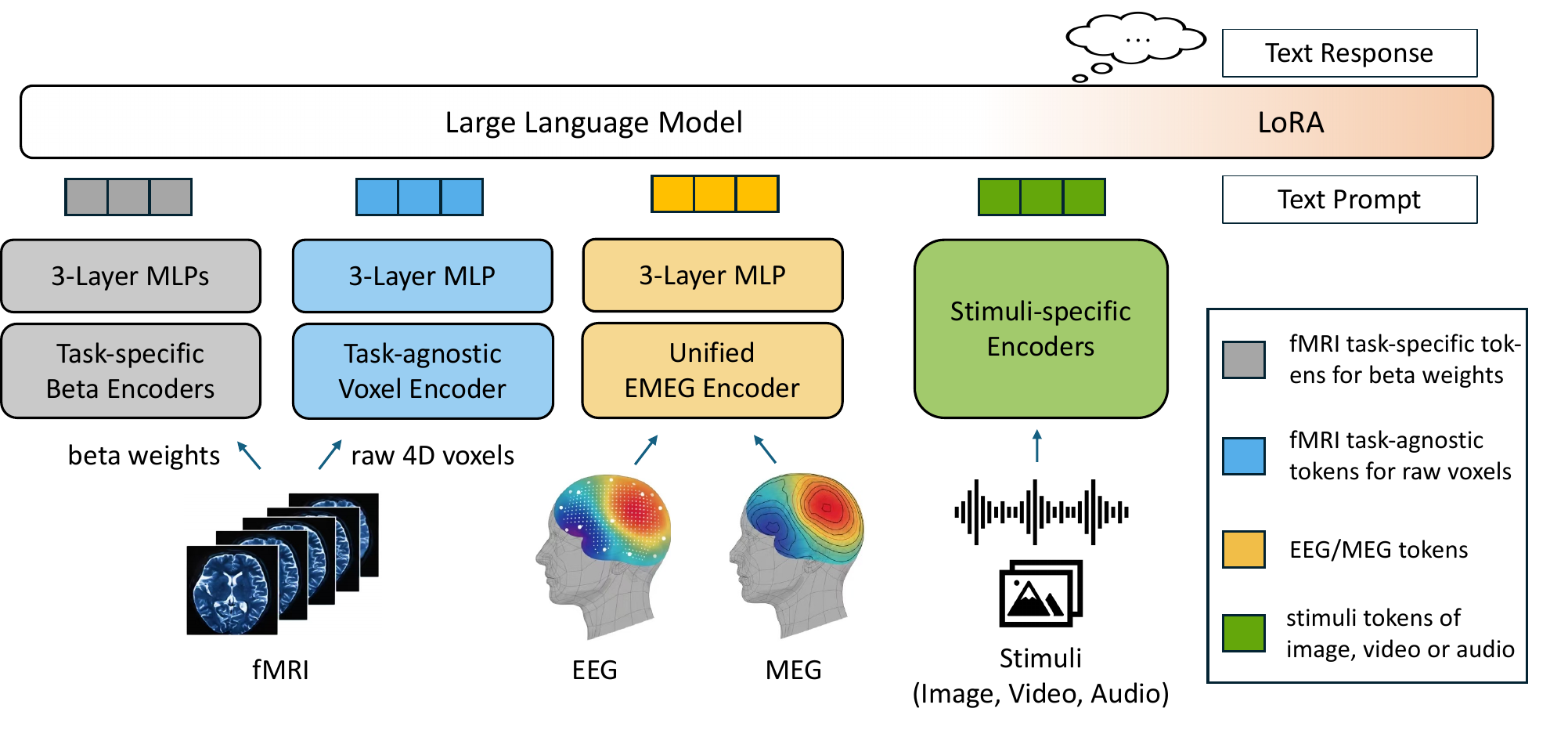}}
\caption{Overview of the NOBEL architecture. The model integrates heterogeneous brain signals and external stimuli into an LLM, features dual-path fMRI encoders for disentangling metabolic features (raw voxels vs. beta weights) and a unified EEG/MEG encoder for electromagnetic dynamics. All brain modalities are projected into the LLM's input space via modality-specific 3-Layer MLP aligners. The system supports flexible input combinations and is optimized via LoRA for instruction-following tasks.}
\label{fig:architecture}
\end{center}
\vspace{-1cm}
\end{figure*}

NOBEL is a unified foundation model designed to decode omni-modal non-invasive brain signals, ranging from low-frequency fMRI activities to high-frequency EEG and MEG signals, within a shared semantic space. As illustrated in Fig.~\ref{fig:architecture}, the framework operates as a general-purpose interpreter, which accepts a flexible combination of fMRI, EEG, MEG, and external stimuli, controlled via user-provided textual prompts to generate natural language responses. To maintain signal granularity and task flexibility, fMRI signals are processed as raw 4D voxels and GLM beta weights.

To bridge the substantial domain gap between biological signals and the LLM's textual embedding space, we design specialized encoding streams that converge via a unified alignment mechanism. For metabolic signals, we implement a dual-path strategy to explicitly disentangle physiological state information from task-related semantic information. 
The first path is \textit{task-agnostic}, designed to capture universal physiological features embedded in raw voxels. We employ a universal encoder $\mathcal{E}_{\text{vox}}$ based on NeuroSTORM~\cite{wang2025towards} to process the raw 4D input $\mathbf{X}_{\text{vox}}$. 
The second path is \textit{task-specific}, aiming to extract high-level semantic content from GLM beta weights. Recognizing that semantic representations vary significantly across different modalities (\textit{e.g.}, viewing an image and watching a video), we instantiate a set of task-specific encoders $\{\mathcal{E}_{\text{beta}}^{(k)}\}_{k=1}^{K}$, where each encoder corresponds to a specific stimulus domain $k$. 
To map these heterogeneous features into the LLM's input space, each encoding branch is followed by a modality aligner $\text{Aligner}(\cdot)$, implemented as a 3-layer multilayer perceptron (MLP) with GELU activation \cite{hendrycks2016gaussian}. Consequently, the unified metabolic tokens $\mathbf{H}_{\text{vox}}$ and the task-specific semantic tokens $\mathbf{H}_{\text{beta}}^{(k)}$ are formally derived as:
\begin{align}
    \mathbf{H}_{\text{vox}} &= \text{Aligner}_{\text{vox}}\big(\mathcal{E}_{\text{vox}}(\mathbf{X}_{\text{vox}})\big),\\
\mathbf{H}_{\text{beta}}^{(k)} &= \text{Aligner}_{\text{beta}}^{(k)}\big(\mathcal{E}_{\text{beta}}^{(k)}(\mathbf{X}_{\text{beta}}^{(k)})\big).
\end{align}
where $\mathbf{X}_{\text{beta}}^{(k)}$ denotes the beta weights associated with the $k$-th task. This design allows NOBEL to simultaneously leverage global physiological context and precise, stimulus-aligned semantic details.

Complementing the metabolic perspective, we model high-frequency electromagnetic signals using a unified EEG/MEG encoder. Although EEG and MEG measure different physical quantities (electrical potentials and magnetic flux, respectively), both arise from the same underlying neuronal current sources. Exploiting this shared biophysical origin, we adopt the BrainOmni backbone $\mathcal{E}_{\text{em}}$ to extract unified temporal–spectral representations from the input sequence $\mathbf{X}_{\text{em}}$. The resulting electromagnetic tokens $\mathbf{H}_{\text{em}}$ are then mapped into the semantic space of the LLM through a modality-specific three-layer MLP aligner as:
\begin{equation}
    \mathbf{H}_{\text{em}} = \text{Aligner}_{\text{em}}\big(\mathcal{E}_{\text{em}}(\mathbf{X}_{\text{em}})\big).
\end{equation}
Furthermore, to ground these abstract neural representations in their perceptual context, the model incorporates external stimuli via a simuli-aware encoding stream. Similar to the task-specific fMRI path, distinct encoders $\mathcal{E}_{\text{stim}}^{(m)}$ are employed for different stimulus modalities $m$ (\textit{e.g.}, visual or audio) to process the raw media inputs $\mathbf{X}_{\text{stim}}^{(m)}$.
The modality-specific stimulus tokens $\mathbf{H}_{\text{stim}}^{(m)}$ are obtained as:
\begin{equation}
    \mathbf{H}_{\text{stim}}^{(m)} = \mathcal{E}_{\text{stim}}^{(m)}(\mathbf{X}_{\text{stim}}^{(m)}).
\end{equation}

The integration framework is inherently flexible, allowing NOBEL to process arbitrary combinations of available inputs. We dynamically construct the LLM-input sequence $\mathbf{S}$ by concatenating the user prompt $\mathbf{T}_{\text{prompt}}$ with the aligned tokens of all present modalities. Let $\mathcal{M}$ and $\mathcal{K}$ denote the sets of available stimulus modalities and task-specific beta inputs, and let $\mathbb{I}_{(\cdot)} \in \{0, 1\}$ denote the indicator function for the presence of the corresponding modality. The final input token sequence to LLM is formulated as:
\begin{equation}\label{equ:cat}
\begin{aligned}
    \mathbf{S} = \mathbf{T}_{\text{prompt}} & \left(\oplus_{m \in \mathcal{M}} \mathbf{H}_{\text{stim}}^{(m)}\right) \oplus (\mathbb{I}_{\text{em}} \cdot \mathbf{H}_{\text{em}}) \\
    & \left(\oplus_{k \in \mathcal{K}} \mathbf{H}_{\text{beta}}^{(k)}\right) \oplus (\mathbb{I}_{\text{vox}} \cdot \mathbf{H}_{\text{vox}}).  
\end{aligned}
\end{equation}
where $\oplus$ denotes the sequence concatenation operator, and terms with $\mathbb{I}=0$ or empty sets are omitted.

To efficiently adapt the pre-trained multimodal LLM to the neuroscientific domain, low-rank adaptation (LoRA) \cite{hu2022lora} is used, where we freeze the LLM's original parameters and inject trainable low-rank matrices into the attention layers. 

\subsection{Training Pipeline}
\label{sec:training_pipeline}

Building a unified model that comprehends heterogeneous brain signals and stimuli requires large-scale multimodal training. 
NOBEL leverages Qwen2.5-Omni \cite{xu2025qwen2} as the backbone, whose visual and auditory encoders are directly inherited and thus inherently aligned with the LLM's semantic space. While this ensures robust handling of external stimuli, the task-specific fMRI branch faces a unique challenge: decoding subtle, stimulus-evoked perceptual information via beta weights is significantly more difficult than extracting global physiological meta-information due to the low SNR. Directly mapping these noisy, high-dimensional features to the LLM space often leads to optimization instability and poor semantic grounding.
To address this, we implement a targeted pre-alignment strategy for the fMRI beta branch prior to the joint multimodal training. Drawing inspiration from MindEye2 \cite{scotti2024mindeye}, we prioritize aligning the noisy neural representations with the robust sensory features already encoded by Qwen2.5-Omni. The output features of the beta encoders and the corresponding frozen stimuli encoders are projected into a shared low-dimensional latent space. A contrastive objective is applied to maximize the similarity between the brain signal embeddings and their corresponding ground-truth stimulus embeddings. To mitigate the limited sample size in fMRI datasets and improve generalization, we additionally adopt MixCo-style neural feature mixing \cite{li2025predictive} during pre-alignment, which interpolates neural representations and their corresponding stimulus embeddings as a form of cross-modal regularization. This process effectively utilizes the high-quality visual/audio representations as ``semantic anchors" to stabilize the fMRI features.
Only after this cross-modal correspondence is established are the pre-aligned features projected into the LLM input space via the MLP aligners.

With the stimulus-related branches pre-aligned, the framework proceeds to unified omni-modal instruction tuning. In this stage, data from all modalities, including EEG, MEG, raw fMRI voxels, fMRI beta weights, and external stimuli, are integrated into the dynamic sequence generation process. All decoding tasks are formulated as brain-to-text generation problems, allowing the model to learn synergistic representations across modalities. The entire system is optimized using the standard autoregressive next token prediction objective. Given the target text response $\mathbf{Y} = \{y_1, y_2, ..., y_L\}$, the loss function is formulated as:
\begin{equation}
    \mathcal{L} = -\sum_{t=1}^{L} \log P(y_t | y_{<t}, \mathbf{S}).
\end{equation}
This training objective ensures that NOBEL learns to generate coherent, context-aware analysis conditioned on the unified omni-modal brain dynamics, while the use of LoRA allows for efficient adaptation of the LLM backbone to the neuroscientific domain.

\section{Experimental Setup}
\label{sec:setup}
\subsection{Data Specifications}
\label{sec:data_spec}


To construct a versatile foundation model capable of decoding the full spectrum of brain dynamics, a diverse corpus of datasets was aggregated to serve as the training substrate. As detailed in Table~\ref{tab:data}, this collection spans high-frequency EEG and MEG signals and low-frequency fMRI maps. While standard clinical datasets such as AD65 and ABIDE-I are included to establish baseline competencies in physiological analysis, specific emphasis is placed on several specialized data categories introduced to endow the model with fine-grained and multimodal abilities.

First, to address the challenge of cross-modal synthesis, scarce paired datasets are incorporated to enable joint learning across different signal types. Specifically, the LEMON \cite{lemon} and Nat-View \cite{natview} datasets are utilized for joint EEG-fMRI processing, while the SMN4Lang \cite{smn4lang} dataset is employed for MEG-fMRI tasks. The SomatoMotor \cite{ds006035:1.0.0} dataset is included to facilitate joint EEG-MEG learning.

Besides, to foster precision signal understanding beyond coarse classification, specialized tasks are designed using the HCP \cite{hcp} dataset by injecting random perturbations ranging from 1\% to 10\% of the fMRI signal amplitude. The model is required to identify the specific perturbed brain regions and regress the corresponding perturbation magnitudes, a process that necessitates the capture of fine-grained variations within the neural dynamics.

To enable the model to capture and interpret the dynamic variations in brain signals induced by external stimuli, the NSD \cite{allen2022massive} and HAD \cite{zhou2023large} datasets are included. Utilizing the beta-encoding stream, these datasets are employed to map specific stimulus-evoked neural patterns to their corresponding high-level semantic content. The training formulation includes generative decoding, where the model translates stimulus-driven brain activity directly into image captions or video classifications, and stimulus-brain verification, which requires distinguishing whether a specific visual input triggered the recorded neural response.
%
Dataset details are included in Appendix~\ref{app:dataset}.

\begin{table*}[ht]
    \centering
    \caption{Overview of the used datasets. The collection spans single-modality clinical datasets, pioneering paired EEG/MEG-fMRI datasets, and complex stimulus-aware semantic decoding datasets. To make a comparison with previous literature, balance accuracy (BAcc) is used to evaluate EEG/MEG classification tasks, while accuracy (Acc) is primarily used to evaluate other classification tasks. We use mean absolute error (MAE) to evaluate HCP stimulus amplitude regression and Rouge-L score to evaluate NSD image captioning.}
    \label{tab:data}
    \resizebox{\textwidth}{!}{
    \setlength{\tabcolsep}{3pt}
    \begin{tabular}{lllc cccc}
    \toprule
    \textbf{Modality} & \textbf{Dataset} & \textbf{Task} &  \textbf{Duration} & \textbf{\#Subject} & \textbf{\#Train Sample} & \textbf{\#Test Sample} & \textbf{Metric $\uparrow$} \\
    \midrule
    \multicolumn{8}{l}{\textit{High-Frequency Electromagnetic Datasets}} \\
    \multirow{8}{*}{EEG} 
    & AD65 & Alzheimer’s Disease Detection & 14.9h & 65 & 3.1k & 1.1k & BAcc\% \\
    & MDD & Depression Detection & 20.3h & 35 & 4.5k & 1.4k & BAcc\% \\
    & PD31 & Parkinson’s Disease Detection & 2.5h & 31 & 0.6k & 0.1k & BAcc\% \\
    & TUAB & Abnormal Signal Detection & 1135.7h & 2328 & 299.1k & 36.9k & BAcc\% \\
    & TUEV & Event Type Classification & 155.9h & 294 & 68.7k & 28.3k & BAcc\% \\
    & FACED & Emotion Recognition & 28.7h & 123 & 6.3k & 2.0k & BAcc\% \\
    & WBCIC\_SHU & Motor Imagery Classification & 34.0h & 51 & 18.6k & 6.0k & BAcc\% \\
    & PhysioNet-MI & Motor Imagery Classification & 10.9h & 109 & 6.1k & 1.9k & BAcc\% \\
    \addlinespace
    \multirow{2}{*}{MEG} 
    & ASD74 & Autism Spectrum Disorder & 34.2h & 74 & 7.8k & 2.2k & BAcc\% \\
    & MEG-MMI & Depression Detection & 14.8h & 51 & 1.1k & 0.3k & BAcc\% \\
    \addlinespace
    \multirow{1}{*}{EEG + MEG} 
    & SomatoMotor & Motor Response Prediction & 0.7h & 5 & 0.7k & 0.2k & BAcc\% \\
    \midrule
    \multicolumn{8}{l}{\textit{Low-Frequency Metabolic Datasets}} \\
    \multirow{10}{*}{\makecell[l]{fMRI \\ (Voxel-based)}}  
    & \multirow{2}{*}{ABIDE-I} & Autism Detection (asd) & \multirow{2}{*}{111.8h} & \multirow{2}{*}{1035} & \multirow{2}{*}{19.7k} & \multirow{2}{*}{2.4k} & Acc\% \\
    & & Gender Classification (sex) &  &  & & & Acc\% \\
    & ADHD200 & ADHD Detection (adhd) & 129.5h & 973 & 19.4k & 3.1k & Acc\% \\
    & \multirow{2}{*}{Neural Processing} & Depression Detection (mdd) & \multirow{2}{*}{21.2h} & \multirow{2}{*}{39} & \multirow{2}{*}{2.9k} & \multirow{2}{*}{0.8k} & Acc\% \\
    & & Gender Classification (sex) &  &  & & & Acc\% \\
    & \multirow{2}{*}{Working Memory} & Schizophrenia Detection (scz) & \multirow{2}{*}{35.2h} & \multirow{2}{*}{99} & \multirow{2}{*}{4.7k} & \multirow{2}{*}{1.5k} & Acc\% \\
    & & Gender Classification (sex) &  &  & &  & Acc\% \\
    & XP1-2 & Gender Classification (sex) & \multirow{1}{*}{13.0h} & 30 & 2.8k & 1.2k & Acc\% \\
    & \multirow{2}{*}{HCP} & Brain Area Classification (area) & \multirow{2}{*}{418.3h} & \multirow{2}{*}{1096} & \multirow{2}{*}{108.2k} & \multirow{2}{*}{9.5k} & Acc\% \\
    & & Stimulus Amplitude Reg. (amp) &  &  &  &  & MAE \\
    \midrule
    \multicolumn{8}{l}{\textit{Synergistic Omni-Modal Datasets}} \\
    \multirow{2}{*}{\makecell[l]{EEG + fMRI \\ (Voxel-based)}}
    & LEMON & Gender Classification & 39.6h & 220 & 6.0k & 0.5k & Acc\% \\
    & Nat-View & Gender Classification & 42.8h & 22 & 8.1k & 1.7k & Acc\% \\
    \addlinespace
    \makecell[l]{MEG + fMRI \\ (Voxel-based)}
    & SMN4Lang & Gender Classification & 70.4h & 12 & 13.6k & 9.6k & Acc\% \\
    \midrule
    \multicolumn{8}{l}{\textit{Stimulus-Aware Datasets}} \\
    \multirow{2}{*}{\makecell[l]{fMRI \\ (Beta-based)}} 
    & NSD & Image Captioning & 284h & 8 & 29.9k & 21.1k & Rouge-L \\
    & HAD & Video Classification & 31.2h & 30 & 19.4k & 2.2k & Acc\% \\
    \addlinespace
    \multirow{2}{*}{\makecell[l]{Stimuli + fMRI \\ (Beta-based)}} 
    & NSD & Image-Brain Verification & 284h & 8 & 59.8k & 42.1k & Acc\% \\
    & HAD & Video-Brain Verification & 31.2h & 30 & 38.8k & 4.3k & Acc\% \\
    \bottomrule
    \end{tabular}
    }
\end{table*}

\subsection{Model and Training Specifications}
The NOBEL framework is instantiated based on Qwen2.5-Omni-7B \cite{xu2025qwen2}. To inherit strong multimodal capabilities, the LLM backbone, as well as the audio and visual stimuli encoders, are initialized with pre-trained weights from Qwen2.5-Omni-7B. For the brain signal branches, we leverage domain-specific foundation models. The unified EMEG encoder is initialized from BrainOmni-base \cite{xiao2025brainomni}, and the fMRI task-agnostic voxel encoder is initialized from NeuroSTORM \cite{wang2025towards}. All remaining components are randomly initialized. 
Regarding the task-specific beta encoding sets $\mathcal{M}$ and $\mathcal{K}$ in Eqn.~\eqref{equ:cat}, our implementation covers two primary categories: static images and dynamic video-audio clips. For the image-evoked beta encoder, we adopt the architecture of MindEye2 \cite{scotti2024mindeye}, while for the video-evoked one, we utilize a lightweight MLP as the encoder.

For pre-alignment for the fMRI beta branch, we use dataset-specific training protocols for NSD and HAD. For the NSD dataset, we use a minibatch size of 128, evenly distributed across 8 subjects. The model is trained with an initial learning rate of $1\times10^{-4}$ and a cosine cyclic scheduler. Training is conducted for 150 epochs. MixCo-based neural feature mixing is applied during the first 50 epochs of pre-alignment only. For the HAD dataset, we use a minibatch size of 32 and an initial learning rate of $1\times10^{-3}$ using a OneCycle scheduler. Pre-alignment on HAD is formulated as a 180-way classification and trained using a cross-entropy loss.


For joint multimodal training, to adapt this massive architecture to the neuroscientific domain efficiently, we employ LoRA with the rank and alpha set to $r=128,\alpha=256$. Optimization is applied to the LoRA, the fMRI encoders, and all modality aligners. Other parameters are kept frozen to preserve their generalizable representations.
The model is first trained with a learning rate of $2 \times 10^{-5}$ and a global batch size of 32 for a total of 90,000 steps with a linear scheduler, then fine-tuned on specialized data. To mitigate the optimization instability caused by the extreme heterogeneity of the input signals, we implement a modality-aware batch sampling strategy, ensuring that all samples within a single training batch belong to the same modality. For inference, we employ a greedy decoding strategy to generate deterministic text responses. The datasets and specific evaluation metrics used are provided in Table~\ref{tab:data}.

\section{Experimental Results}
\label{sec:expres}
\begin{table*}[ht]
    \setlength{\tabcolsep}{3pt}
    \caption{Quantitative results on EEG and MEG tasks of or our model and specialized discriminative baselines, including CNN-Transformer \cite{peh2022transformer}, ContraWR \cite{yang2023self}, SPaRCNet \cite{jing2023development}, ST-Transformer \cite{song2021transformer}, LaBraM \cite{jiang2024labram}, CBraMod \cite{wang2025cbramod}, FAMED \cite{hirano2024deep}, and BrainOmni \cite{xiao2025brainomni}.} 
    \centering
    \resizebox{\textwidth}{!}{
    \begin{tabular}{lccccccccccc}
    \toprule
    \multirow{2}{*}{\textbf{Model}} & \multicolumn{8}{c}{EEG} & \multicolumn{2}{c}{MEG} & \multicolumn{1}{c}{EEG + MEG}\\
    \cmidrule(lr){2-9}  \cmidrule(lr){10-11} \cmidrule(lr){12-12} & AD65 & MDD & PD31 & TUAB & TUEV & FACED & WBCIC\_SHU & PhysioNet-MI & ASD74 & MEG-MMI & SomatoMotor \\
    \midrule
    CNN-Transformer & 77.7 & 86.9 & 50.0 & 80.7 & 32.3 & 36.8 & 65.8 & 32.9 & 50.0 & 50.0 & 50.0 \\
    ContraWR & 78.8 & 81.9 & 45.8 & 81.2 & 41.3 & 33.6 & 59.3 & 36.4 & 50.3 & 52.8 & 50.0 \\
    SPaRCNet & 60.3 & 79.9 & 40.4 & 78.5 & 40.8 & 37.2 & 73.4 & 45.8 & 51.1 & 57.3 & 53.7 \\
    ST-Transformer & 61.2 & 79.6 & 50.0 & 79.5 & 40.7 & 39.4 & 75.5 & 39.9 & 60.5 & \textbf{57.4} & 66.2 \\
    LaBraM & 62.9 & 86.1 & 50.0 & 81.2 & 60.5 & 45.7 & 80.4 & 53.7 & - & - & - \\
    CBraMod & 72.3 & 84.9 & 73.7 & 81.3 & 57.2 & 42.0 & 79.7 & 57.8 & - & - & - \\
    FAMED & - & - & - & - & - & - & - & - & 51.3 & 50.0 & 50.0 \\
    BrainOmni & 80.3 & \textbf{89.1} & \textbf{82.5} & 81.7 & \textbf{64.5} & \textbf{49.7} & \textbf{82.1} & \textbf{58.4} & \textbf{65.3} & 55.8 & \textbf{92.1} \\
    \midrule
    NOBEL & \textbf{81.3} & 85.9 & 74.5 & \textbf{82.4} & 61.7 & 41.2 & 79.4 & 49.1 & 63.6 & 54.2 & 81.3 \\
    \bottomrule
    \end{tabular}
    }
    \label{tab:emegres}
\end{table*}

\begin{table*}[ht]
    \caption{Quantitative results on voxel-based fMRI tasks of our model and other fMRI encoder baselines, including BrainGNN \cite{li2021braingnn}, BrainHarmonix-F \cite{dong2025brain}, SWiFT \cite{kim2023swift}, and NeuroSTORM \cite{wang2025towards}. Baselines marked with (*) are from \citet{dong2025brain}, which may have different data splits from our settings and are included only for reference. NOBEL demonstrates competitive performance in phenotype prediction and retains precise low-level signal perception on HCP.}
    \centering
    \footnotesize
    \begin{tabular}{lcccccccccc}
    \toprule
    \multirow{2}{*}{\textbf{Model}} & \multicolumn{2}{c}{ABIDE-I} & ADHD200 & \multicolumn{2}{c}{Neural Processing} & \multicolumn{2}{c}{Working Memory} & XP1-2  & \multicolumn{2}{c}{HCP}  \\
    \cmidrule(lr){2-3} \cmidrule(lr){4-4} \cmidrule(lr){5-6}  \cmidrule(lr){7-8} \cmidrule(lr){9-9} \cmidrule(lr){10-11} & asd & sex & adhd & mdd & sex & scz & sex & sex  & area & amp  \\
    \midrule
    BrainGNN$^*$ & 56.7 & - & 62.2 & - & - & - & - & - & - & - \\
    BrainHarmonix-F$^*$ & 57.4 & - & \textbf{67.7} & - & - & - & - & - & - & -  \\
    SWiFT & 49.1 & 86.7 & 60.8 & 50.0 & 75.0 & 75.3 & 76.4 & 70.3 & 36.8 & 0.48 \\
    NeuroSTORM & 57.1 & 85.4 & 56.1 & \textbf{73.7} & 75.0 & \textbf{80.6} & \textbf{80.5} & 66.9 & 100 & 0.04 \\
    \midrule
    NOBEL & \textbf{57.6} & \textbf{89.2} & 56.9 & 61.7 & \textbf{83.6} & 72.1 & 79.1 & \textbf{78.8} & \textbf{100} & \textbf{0.01} \\
    \bottomrule
    \end{tabular}
    \vspace{-0.2cm}
    \label{tab:fmrires}
\end{table*}

\subsection{Results on EEG, MEG, or voxel-based fMRI Tasks}
\label{sec:exp_unimodal}

Table~\ref{tab:emegres} and Table~\ref{tab:fmrires} present the quantitative evaluation of NOBEL on unimodal EEG, MEG, and fMRI tasks. When interpreting these results, it is crucial to acknowledge that conventional baselines are typically specialized models, trained and optimized exclusively on single datasets for isolated tasks, while NOBEL is architected as a unified, omni-modal foundation model designed to integrate heterogeneous signals within a shared semantic space. 

Despite this architectural difference, NOBEL demonstrates strong competitiveness. For EEG and MEG tasks, as shown in Table~\ref{tab:emegres}, it consistently outperforms established discriminative baselines such as ST-Transformer and SPaRCNet across the majority of EEG and MEG benchmarks. Furthermore, even when compared against the current state-of-the-art BrainOmni, NOBEL still exhibits competitive performance. Notably, on AD65 and TUAB, NOBEL achieves accuracies of 81.3\% and 82.4\%, respectively, surpassing BrainOmni. This suggests the potential of our generative framework for aligning high-frequency electromagnetic features with the rich semantic space of an LLM.

In the voxel-based fMRI domain, as detailed in Table~\ref{tab:fmrires}, NOBEL exhibits a similar pattern of resilience and capability. Specifically, on gender classification tasks across the ABIDE-I, Neural Processing, and XP1-2 datasets, NOBEL achieves accuracies of 89.2\%, 83.6\%, and 78.8\%, respectively, consistently surpassing both the NeuroSTORM and SWiFT baselines. This indicates that the unified framework effectively captures global phenotypic traits, potentially benefiting from the regularization provided by the LLM's semantic space to generalize better across subjects. Regarding pathology detection tasks such as autism, ADHD, depression, and schizophrenia, NOBEL also delivers competitive performance, maintaining robustness comparable to specialized encoders.
Notably, on the HCP benchmark focusing on perturbed brain area classification and stimulus amplitude regression, NOBEL matches the high precision of the NeuroSTORM encoder, which demonstrates that our model successfully retains the capabilities of the underlying encoder. This confirms that the alignment process preserves the precision required to perceive subtle, low-level information in fMRI, without losing granularity during the projection to the semantic space.

\subsection{Results on stimuli-aware beta-based fMRI tasks}
\begin{table}[ht]
    \caption{Results on stimulus-aware semantic tasks.
    We compare (i) fMRI-only decoding (classification for HAD and captioning for NSD) and (ii) stimuli-conditioned classification (verifying whether the given stimuli triggered fMRI). Note that the MindEye2$*$ baseline is reported on a single-subject split, whereas NOBEL is evaluated in a more challenging multi-subject setting.
    }
    \centering
    \footnotesize
    \begin{tabular}{lcccc}
    \toprule
    \multirow{2}{*}{\textbf{Model}} & \multicolumn{2}{c}{fMRI only} & \multicolumn{2}{c}{fMRI + stimuli}\\
    \cmidrule(lr){2-3} \cmidrule(lr){4-5} & HAD & NSD &  HAD & NSD \\
    \midrule
    MLP baseline & 78.0 & - & - & - \\
    MindEye2$^*$ & - & 32.6 & - & - \\
    NOBEL & 74.6 & 29.7 & 62.8 & 92.7 \\
    \bottomrule
    \end{tabular}
    \vspace{-0.5cm}
    \label{tab:betares}
\end{table}

Table~\ref{tab:betares} summarizes the performance on stimulus-aware tasks using the HAD and NSD datasets. These tasks are categorized into two settings: fMRI only decoding, and stimuli-input cross-modal verification.

In the fMRI only decoding setting, the model interprets semantic content solely from brain signals. For the HAD video classification task, NOBEL achieves an accuracy of 74.6\%, which is competitive with the specialized MLP baseline. On the more challenging NSD image captioning task, NOBEL demonstrates robust semantic understanding with a Rouge-L score of 29.7 across all subjects. It is important to note that the MindEye2 baseline is only for a single-subject split. In contrast, NOBEL handles a multi-subject setting simultaneously. Although evaluating on heterogeneous functional topography across multiple subjects is more challenging than modeling a single subject, our model still achieves high linguistic correspondence. Appendix~\ref{app:case} visualizes qualitative examples of these decoded captions.

For the stimuli-input cross-modal verification setting, which involves processing both external stimuli and brain signals, NOBEL demonstrates unique capabilities. Since stimuli and fMRI reside in distinct modalities, existing baselines struggle to process them concurrently. However, by aligning omni-modal signals within the semantic space of the LLM, NOBEL effectively analyzes the relationship between the sensory input and the neural response. In this task, the model determines whether a specific image or video triggered the recorded fMRI signal. NOBEL achieves an accuracy of 62.8\% on HAD and reaches 92.7\% on NSD. The high accuracy on NSD indicates that the model has successfully learned a precise alignment between visual features and their corresponding cortical representations, enabling it to verify the causal link between stimulus and brain activity.

\vspace{-0.2cm}
\subsection{Synergistic Omni-Modal Decoding}
\label{sec:exp_multimodal}

The most significant distinction between NOBEL and existing brain foundation models lies in its capability to synthesize heterogeneous signals. While previous works are confined to isolated modalities, NOBEL bridges the ``frequency gap", integrating electromagnetic signals with metabolic maps. To validate this synergistic capability, we evaluate the model on paired datasets where cross-modal recordings are available. Table~\ref{tab:emegfmrires} presents the comparative results.

\begin{table}[ht]
    \caption{Results on synergistic cross-modal tasks. We compare the model trained on isolated modalities versus joint inputs. Results show that synthesizing EEG-MEG and fMRI signals consistently yields superior performance compared to using either alone.}
    \centering
    \resizebox{\linewidth}{!}{
    \begin{tabular}{lccc}
    \toprule
    \multirow{2}{*}{\textbf{Model}} & \multicolumn{2}{c}{EEG + fMRI} & \multicolumn{1}{c}{MEG + fMRI} \\
    & LEMON & Nat-View & SMN4Lang \\
    \midrule
    EEG-MEG-only baseline & 70.6 & 16.7 & 77.4 \\
    fMRI-only baseline & 76.1 & 86.6 & 69.0 \\
    NOBEL (EEG-MEG + fMRI) & \textbf{81.8} & \textbf{92.8} & \textbf{78.5} \\
    \bottomrule
    \end{tabular}
    }
    \vspace{-0.2cm}
    \label{tab:emegfmrires}
\end{table}

Across all benchmarks, NOBEL consistently outperforms the best-performing single-modality baselines. On LEMON and SMN4Lang, the joint model leverages the complementary nature of the signals to boost accuracy beyond what is achievable by either EEG-MEG or fMRI alone. This trend is even more pronounced on the Nat-View dataset, where the EEG-only branch yields below-chance performance (16.7\%), likely because the data relies heavily on spatial features that scalp-EEG fails to capture. Rather than being hampered by the weaker modality, the joint model utilizes the fMRI information while extracting subtle temporal cues from the EEG, pushing the final accuracy to 92.8\%. These results validate that NOBEL does not merely average the outputs of separate encoders. Instead, the LLM backbone acts as a semantic integrator, effectively aligning the millisecond-level dynamics of EEG-MEG with the voxel-level topology of fMRI within a shared semantic space.

\vspace{-0.2cm}
\section{Conclusion}
\label{sec:conclude}
This paper introduces NOBEL, the first foundation LLM designed to unify non-invasive high-frequency EEG-MEG signals and low-frequency fMRI signals within a generative LLM framework. By orchestrating a unified EEG-MEG encoder and a novel dual-path fMRI strategy, NOBEL supports the comprehensive processing of EEG, MEG, and fMRI, while simultaneously integrating external audio-visual stimuli into a shared semantic space. Our extensive experimental evaluation confirms that NOBEL operates as a robust generalist: it not only retains the discriminative capacity of single-modal encoders across standard physiological tasks but also unlocks cross-modal capabilities unattainable by isolated models. Compelling empirical evidence is provided that the synergistic fusion of EEG-MEG and fMRI yields superior decoding accuracy through cross-modal complementarity, outperforming unimodal baselines. The model also excels in stimulus-aware tasks, achieving robust performance in decoding visual semantics from multi-subject fMRI data on NSD and HAD. Uniquely, the incorporation of external stimuli as inputs endows the model with semantic reasoning powers, enabling it to verify the causal link between sensory inputs and neural responses. By processing EEG-MEG and fMRI brain dynamics in a single model, NOBEL takes a step towards unifying non-invasive brain decoding, showing the potential of omni-modal brain understanding.

\section*{Impact Statement}

This paper presents work whose goal is to advance the field of Machine Learning. There are many potential societal consequences of our work, none which we feel must be specifically highlighted here.




\nocite{langley00}

\bibliography{example_paper}

@inproceedings{xiao2025brainomni,
    title={{BrainOmni: A Brain Foundation Model for Unified {EEG} and {MEG} Signals}},
    author={Qinfan Xiao and Ziyun Cui and Chi Zhang and SiQi Chen and Wen Wu and Andrew Thwaites and Alexandra Woolgar and Bowen Zhou and Chao Zhang},
    booktitle={Proc. NeurIPS},
    year={2025}
}

@article{wang2025towards,
  title={{Towards a General-purpose Foundation Model for fMRI Analysis}},
  author={Wang, Cheng and Jiang, Yu and Peng, Zhihao and Li, Chenxin and Bang, Changbae and Zhao, Lin and Lv, Jinglei and Sepulcre, Jorge and Yang, Carl and He, Lifang and others},
  journal={arXiv preprint arXiv:2506.11167},
  year={2025}
}

@inproceedings{scotti2024mindeye,
    title={{MindEye2: Shared-Subject Models Enable f{MRI}-To-Image With 1 Hour of Data}},
    author={Paul Steven Scotti and Mihir Tripathy and Cesar Torrico and Reese Kneeland and Tong Chen and Ashutosh Narang and Charan Santhirasegaran and Jonathan Xu and Thomas Naselaris and Kenneth A. Norman and Tanishq Mathew Abraham},
    booktitle={Proc. ICML},
    year={2024},
}

@inproceedings{peh2022transformer,
  title={{Transformer Convolutional Neural Networks for Automated Artifact Detection in Scalp EEG}},
  author={Peh, Wei Yan and Yao, Yuanyuan and Dauwels, Justin},
  booktitle={Proc. EMBC},
  year={2022},
}

@article{song2021transformer,
  title={{Transformer-based Spatial-Temporal Feature Learning for EEG Decoding}},
  author={Song, Yonghao and Jia, Xueyu and Yang, Lie and Xie, Longhan},
  journal={arXiv preprint arXiv:2106.11170},
  year={2021}
}

@article{jing2023development,
  title={{Development of Expert-level Classification of Seizures and Rhythmic and Periodic Patterns During EEG Interpretation}},
  author={Jing, Jin and Ge, Wendong and Hong, Shenda and Fernandes, Marta Bento and Lin, Zhen and Yang, Chaoqi and An, Sungtae and Struck, Aaron F and Herlopian, Aline and Karakis, Ioannis and others},
  journal={Neurology},
  year={2023},
  publisher={Lippincott Williams \& Wilkins Hagerstown, MD}
}

@article{hirano2024deep,
  title={{Deep Learning Based Automatic Detection and Dipole Estimation of Epileptic Discharges in MEG: a Multi-Center Study}},
  author={Hirano, Ryoji and Asai, Miyako and Nakasato, Nobukazu and Kanno, Akitake and Uda, Takehiro and Tsuyuguchi, Naohiro and Yoshimura, Masaki and Shigihara, Yoshihito and Okada, Toyoji and Hirata, Masayuki},
  journal={Scientific Reports},
  year={2024},
  publisher={Nature Publishing Group UK London}
}

@article{yang2023self,
  title={{Self-supervised Electroencephalogram Representation Learning for Automatic Sleep Staging: Model Development and Evaluation Study}},
  author={Yang, Chaoqi and Xiao, Cao and Westover, M Brandon and Sun, Jimeng},
  journal={JMIR AI},
  year={2023},
  publisher={JMIR Publications Inc., Toronto, Canada}
}

@inproceedings{jiang2024labram,
  title={{Large Brain Model for Learning Generic Representations with Tremendous EEG Data in BCI}},
  author={Jiang, Wei-Bang and Zhao, Li-Ming and Lu, Bao-Liang},
  booktitle={Proc. ICLR},
  year={2024}
}

@inproceedings{wang2025cbramod,
  title={{CBraMod: A Criss-Cross Brain Foundation Model for EEG Decoding}},
  author={Wang, Jiquan and Zhao, Sha and Luo, Zhiling and others},
  booktitle={Proc. ICLR},
  year={2025}
}

@article{ouahidi2025reve,
  title={{REVE: A Foundation Model for EEG--Adapting to Any Setup with Large-Scale Pretraining on 25,000 Subjects}},
  author={Ouahidi, Yassine El and Lys, Jonathan and Th{\"o}lke, Philipp and Farrugia, Nicolas and Pasdeloup, Bastien and Gripon, Vincent and Jerbi, Karim and Lioi, Giulia},
  journal={arXiv preprint arXiv:2510.21585},
  year={2025}
}

@inproceedings{yi2023learning,
title={{Learning Topology-Agnostic {EEG} Representations with Geometry-Aware Modeling}},
author={Ke Yi and Yansen Wang and Kan Ren and Dongsheng Li},
booktitle={Proc. NeurIPS},
year={2023},
}

@article{kawahara2017brainnetcnn,
  title={{BrainNetCNN: Convolutional Neural Networks for Brain Networks; Towards Predicting Neurodevelopment}},
  author={Kawahara, Jeremy and Brown, Colin J and Miller, Steven P and Booth, Brian G and Chau, Vann and Grunau, Ruth E and Zwicker, Jill G and Hamarneh, Ghassan},
  journal={NeuroImage},
  year={2017},
  publisher={Elsevier}
}

@article{li2021braingnn,
  title={{BrainGNN: Interpretable Brain Graph Neural Network for fMRI Analysis}},
  author={Li, Xiaoxiao and Zhou, Yuan and Dvornek, Nicha and Zhang, Muhan and Gao, Siyuan and Zhuang, Juntang and Scheinost, Dustin and Staib, Lawrence H and Ventola, Pamela and Duncan, James S},
  journal={Medical Image Analysis},
  year={2021},
  publisher={Elsevier}
}

@inproceedings{kan2022brain,
  title={{Brain Network Transformer}},
  author={Kan, Xuan and Dai, Wei and Cui, Hejie and Zhang, Zilong and Guo, Ying and Yang, Carl},
  booktitle={Proc. NeurIPS},
  year={2022}
}

@inproceedings{kan2022fbnetgen,
  title={{FBNETGEN: Task-aware GNN-based fMRI Analysis via Functional Brain Network Generation}},
  author={Kan, Xuan and Cui, Hejie and Lukemire, Joshua and Guo, Ying and Yang, Carl},
  booktitle={Proc. MIDL},
  year={2022},
}

@article{wei2025fmri,
  title={{fMRI-LM: Towards a Universal Foundation Model for Language-Aligned fMRI Understanding}},
  author={Wei, Yuxiang and Zhang, Yanteng and Xiao, Xi and Qian, Chengxuan and Wang, Tianyang and Calhoun, Vince D},
  journal={arXiv preprint arXiv:2511.21760},
  year={2025}
}

@article{dong2025brain,
  title={{Brain Harmony: A Multimodal Foundation Model Unifying Morphology and Function into 1D Tokens}},
  author={Dong, Zijian and Li, Ruilin and Chong, Joanna Su Xian and Dehestani, Niousha and Teng, Yinghui and Lin, Yi and Li, Zhizhou and Zhang, Yichi and Xie, Yapei and Ooi, Leon Qi Rong and others},
  journal={arXiv preprint arXiv:2509.24693},
  year={2025}
}

@article{kim2023swift,
  title={{Swift: Swin 4D fMRI Transformer}},
  author={Kim, Peter and Kwon, Junbeom and Joo, Sunghwan and Bae, Sangyoon and Lee, Donggyu and Jung, Yoonho and Yoo, Shinjae and Cha, Jiook and Moon, Taesup},
  journal={Proc. NeurIPS},
  year={2023}
}

@article{ozcelik2023natural,
  title={{Natural Scene Reconstruction from fMRI Signals Using Generative Latent Diffusion}},
  author={Ozcelik, Furkan and VanRullen, Rufin},
  journal={Scientific Reports},
  year={2023},
  publisher={Nature Publishing Group UK London}
}

@article{scotti2023reconstructing,
  title={{Reconstructing The Mind's Eye: fMRI-to-image with Contrastive Learning and Diffusion Priors}},
  author={Scotti, Paul and Banerjee, Atmadeep and Goode, Jimmie and Shabalin, Stepan and Nguyen, Alex and Dempster, Aidan and Verlinde, Nathalie and Yundler, Elad and Weisberg, David and Norman, Kenneth and others},
  journal={Proc. NeurIPS},
  year={2023}
}

@article{chen2025mindgpt,
  title={{MindGPT: Interpreting What You See With Non-invasive Brain Recordings}},
  author={Chen, Jiaxuan and Qi, Yu and Wang, Yueming and Pan, Gang},
  journal={IEEE Transactions on Image Processing},
  year={2025},
  publisher={IEEE}
}

@article{tang2023semantic,
  title={{Semantic Reconstruction of Continuous Language from Non-invasive Brain Recordings}},
  author={Tang, Jerry and LeBel, Amanda and Jain, Shailee and Huth, Alexander G},
  journal={Nature Neuroscience},
  year={2023},
  publisher={Nature Publishing Group US New York}
}

@article{horikawa2025mind,
  title={{Mind Captioning: Evolving Descriptive Text of Mental Content from Human Brain Activity}},
  author={Horikawa, Tomoyasu},
  journal={Science Advances},
  year={2025},
  publisher={American Association for the Advancement of Science}
}

@article{fu2025making,
  title={{Making Your Dreams A Reality: Decoding the Dreams into a Coherent Video Story from fMRI Signals}},
  author={Fu, Yanwei and Gao, Jianxiong and Yang, Baofeng and Feng, Jianfeng},
  journal={arXiv preprint arXiv:2501.09350},
  year={2025}
}

@inproceedings{takagi2023high,
  title={{High-resolution Image Reconstruction with Latent Diffusion Models from Human Brain Activity}},
  author={Takagi, Yu and Nishimoto, Shinji},
  booktitle={Proc. CVPR},
  year={2023}
}

@inproceedings{radford2021learning,
  title={{Learning Transferable Visual Models from Natural Language Supervision}},
  author={Radford, Alec and Kim, Jong Wook and Hallacy, Chris and Ramesh, Aditya and Goh, Gabriel and Agarwal, Sandhini and Sastry, Girish and Askell, Amanda and Mishkin, Pamela and Clark, Jack and others},
  booktitle={Proc. ICML},
  year={2021},
}

@article{hendrycks2016gaussian,
  title={{Gaussian Error Linear Units} ({GELU}s)},
  author={Hendrycks, Dan and Gimpel, Kevin},
  journal={arXiv preprint arXiv:1606.08415},
  year={2016}
}

@inproceedings{hu2022lora,
  title={{LoRA: Low-rank Adaptation of Large Language Models.}},
  author={Hu, Edward J and Shen, Yelong and Wallis, Phillip and Allen-Zhu, Zeyuan and Li, Yuanzhi and Wang, Shean and Wang, Lu and Chen, Weizhu and others},
  booktitle={Proc. ICLR},
  year={2022}
}

@article{xu2025qwen2,
  title={{Qwen2.5-Omni Technical Report}},
  author={Xu, Jin and Guo, Zhifang and He, Jinzheng and Hu, Hangrui and He, Ting and Bai, Shuai and Chen, Keqin and Wang, Jialin and Fan, Yang and Dang, Kai and others},
  journal={arXiv preprint arXiv:2503.20215},
  year={2025}
}

@article{allen2022massive,
  title={{A Massive 7T fMRI Dataset to Bridge Cognitive Neuroscience and Artificial Intelligence}},
  author={Allen, Emily J and St-Yves, Ghislain and Wu, Yihan and Breedlove, Jesse L and Prince, Jacob S and Dowdle, Logan T and Nau, Matthias and Caron, Brad and Pestilli, Franco and Charest, Ian and others},
  journal={Nature neuroscience},
  year={2022},
  publisher={Nature Publishing Group US New York}
}

@article{zhou2023large,
  title={{A Large-scale fMRI Dataset for Human Action Recognition}},
  author={Zhou, Ming and Gong, Zhengxin and Dai, Yuxuan and Wen, Yushan and Liu, Youyi and Zhen, Zonglei},
  journal={Scientific Data},
  year={2023},
  publisher={Nature Publishing Group UK London}
}

@article{li2025predictive,
  title={{Predictive Vision-language Integration in The Human Visual Cortex}},
  author={Li, Shurui and Jin, Zheyu and Zhang, Ru-Yuan and Gu, Shi and Li, Yuanning},
  journal={bioRxiv},
  year={2025},
  publisher={Cold Spring Harbor Laboratory}
}

@article{lemon,
  title={{A Mind-brain-body Dataset of MRI, EEG, Cognition, Emotion, and Peripheral Physiology in Young and Old Adults}},
  author={Babayan, Anahit and Erbey, Miray and Kumral, Deniz and Reinelt, Janis D and Reiter, Andrea MF and R{\"o}bbig, Josefin and Schaare, H Lina and Uhlig, Marie and Anwander, Alfred and Bazin, Pierre-Louis and others},
  journal={Scientific data},
  year={2019},
  publisher={Nature Publishing Group}
}

@article{natview,
  title={{An Open-access Dataset of Naturalistic Viewing Using Simultaneous EEG-fMRI}},
  author={Telesford, Qawi K and Gonzalez-Moreira, Eduardo and Xu, Ting and Tian, Yiwen and Colcombe, Stanley J and Cloud, Jessica and Russ, Brian E and Falchier, Arnaud and Nentwich, Maximilian and Madsen, Jens and others},
  journal={Scientific Data},
  year={2023},
  publisher={Nature Publishing Group UK London}
}

@article{smn4lang,
  title={{A Synchronized Multimodal Neuroimaging Dataset for Studying Brain Language Processing}},
  author={Wang, Shaonan and Zhang, Xiaohan and Zhang, Jiajun and Zong, Chengqing},
  journal={Scientific Data},
  year={2022},
  publisher={Nature Publishing Group UK London}
}

@article{hcp,
  title={{The WU-Minn Human Connectome Project: An Overview}},
  author={Van Essen, David C and Smith, Stephen M and Barch, Deanna M and Behrens, Timothy EJ and Yacoub, Essa and Ugurbil, Kamil and Wu-Minn HCP Consortium and others},
  journal={Neuroimage},
  year={2013},
  publisher={Elsevier}
}

@article{TUAB2016,
    title = {{The Temple University Hospital {EEG} Data Corpus}},
    author = {Iyad Obeid and Joseph Picone},
    journal = {Frontiers in Neuroscience},
    year = {2016},
    howpublished={doi: 10.3389/fnins.2016.00196},
}

@misc{MDD2016,
    author = {Mumtaz, Wajid},
    title = {{MDD Patients and Healthy Controls EEG Data (New)}},
    year = {2016},
    howpublished={doi: 10.6084/m9.figshare.4244171.v2},
    url = {https://doi.org/10.6084/m9.figshare.4244171.v2}
}

@article{Yang2025,
  author = {Yang, Banghua and Rong, Fenqi and Xie, Yunlong and Li, Du and Zhang, Jiayang and Li, Fu and Shi, Guangming and Gao, Xiaorong},
  title = {{A Multi-day and High-quality EEG Dataset for Motor Imagery Brain-computer Interface}},
  journal = {Scientific Data},
  year = {2025}
}

@article{Schalk2004,
  author = {Schalk, Gerwin and McFarland, Dennis J. and Hinterberger, Thilo and Birbaumer, Niels and Wolpaw, Jonathan R.},
  title = {{BCI2000: A General-purpose Brain-computer Interface (BCI) system}},
  journal = {IEEE Transactions on Biomedical Engineering},
  year = {2004}
}

@article{Chen2023,
  author = {Chen, Jingjing and Wang, Xiaobin and Huang, Chen and Hu, Xin and Shen, Xinke and Zhang, Dan},
  title = {{A Large Finer-grained Affective Computing EEG Dataset}},
  journal = {Scientific Data},
  year = {2023}
}

@misc{ds002778:1.0.5,
  author = {Alexander P. Rockhill and Nicko Jackson and Jobi George and Adam Aron and Nicole C. Swann},
  title = {{UC} {S}an {D}iego {R}esting {S}tate {EEG} {D}ata from {P}atients with {P}arkinson's {D}isease},
  year = {2021},
  howpublished={doi:10.18112/openneuro.ds002778.v1.0.5},
  publisher = {OpenNeuro}
}

@misc{ds005234:2.1.7,
  author = {Kirill A. Fadeev and Ilacai V. Romero Reyes and Dzerassa D. Goiaeva and Tatyana S. Obukhova and Tatyana M. Ovsiannikova and Andrey O. Prokofyev and Anna M. Rytikova and Artyom Y. Novikov and Vladimir V. Kozunov and Tatyana A. Stroganova and Elena V. Orekhova},
  title = {{Perception of vowel sounds in children with autism spectrum disorders and typically developing children ({MEG/ERF} study}},
  year = {2024},
  howpublished={doi:10.18112/openneuro.ds005234.v2.1.7},
  publisher = {OpenNeuro}
}

@misc{ds006035:1.0.0,
  author = {Fa-Hsuan Lin and Deirdre Foxe von Pechmann and Kaisu Lankinen and Seppo Ahlfors and Bruce Rosen and Jyrki Ahveninen and Matti Hämäläinen and Tommi Raij},
  title = {{Somatomotor}},
  year = {2025},
  howpublished={doi:10.18112/openneuro.ds006035.v1.0.0},
  publisher = {OpenNeuro}
}

@misc{ds004504:1.0.2,
  author = {Andreas Miltiadous and Katerina D. Tzimourta and Theodora Afrantou and Panagiotis Ioannidis and Nikolaos Grigoriadis and Dimitrios G. Tsalikakis and Pantelis Angelidis and Markos G. Tsipouras and Evripidis Glavas and Nikolaos Giannakeas and Alexandros T. Tzallas},
  title = {{A dataset of 88 {EEG} recordings from: {A}lzheimer's disease, {F}rontotemporal dementia and Healthy subjects}},
  year = {2023},
  howpublished={doi:10.18112/openneuro.ds004504.v1.0.2},
  publisher = {OpenNeuro}
}

@misc{ds003568:1.0.4,
  author = {Lucrezia Liuzzi AND Katharine Chang AND Hanna Keren AND Charles Zheng AND Dipta Saha AND Dylan Nielson AND Argyris Stringaris},
  title = {{Mood Induction in MDD and Healthy Adolescents}},
  year = {2023},
  howpublished = {doi:10.18112/openneuro.ds003568.v1.0.4},
  publisher = {OpenNeuro}
}

@article{esteban2019fmriprep,
  title={{fMRIPrep: A Robust Preprocessing Pipeline for Functional MRI}},
  author={Esteban, Oscar and Markiewicz, Christopher J and Blair, Ross W and Moodie, Craig A and Isik, A Ilkay and Erramuzpe, Asier and Kent, James D and Goncalves, Mathias and DuPre, Elizabeth and Snyder, Madeleine and others},
  journal={Nature methods},
  year={2019},
  publisher={Nature Publishing Group US New York}
}

@article{craddock2013neuro,
  title={{The Neuro Bureau Preprocessing Initiative: Open Sharing of Preprocessed Neuroimaging Data and Derivatives}},
  author={Craddock, Cameron and Benhajali, Yassine and Chu, Carlton and Chouinard, Francois and Evans, Alan and Jakab, Andr{\'a}s and Khundrakpam, Budhachandra Singh and Lewis, John David and Li, Qingyang and Milham, Michael and others},
  journal={Frontiers in Neuroinformatics},
  year={2013}
}

@article{bellec2017neuro,
  title={{The Neuro Bureau ADHD-200 Preprocessed Repository}},
  author={Bellec, Pierre and Chu, Carlton and Chouinard-Decorte, Francois and Benhajali, Yassine and Margulies, Daniel S and Craddock, R Cameron},
  journal={Neuroimage},
  year={2017},
  publisher={Elsevier}
}

@article{lepping2016neural,
  title={{Neural Processing of Emotional Musical and Nonmusical Stimuli in Depression}},
  author={Lepping, Rebecca J and Atchley, Ruth Ann and Chrysikou, Evangelia and Martin, Laura E and Clair, Alicia A and Ingram, Rick E and Simmons, W Kyle and Savage, Cary R},
  journal={PloS one},
  year={2016},
  publisher={Public Library of Science San Francisco, CA USA}
}

@article{repovvs2012working,
  title={{Working Memory Related Brain Network Connectivity in Individuals with Schizophrenia and Their Siblings}},
  author={Repov{\v{s}}, Grega and Barch, Deanna M},
  journal={Frontiers in human neuroscience},
  year={2012},
  publisher={Frontiers Research Foundation}
}

@article{lioi2020simultaneous,
  title={{Simultaneous EEG-fMRI During a Neurofeedback Task, a Brain Imaging Dataset for Multimodal Data Integration}},
  author={Lioi, Giulia and Cury, Claire and Perronnet, Lorraine and Mano, Marsel and Bannier, Elise and L{\'e}cuyer, Anatole and Barillot, Christian},
  journal={Scientific data},
  year={2020},
  publisher={Nature Publishing Group UK London}
}

@article{yang2024neurobind,
  title={{NeuroBind: Towards Unified Multimodal Representations for Neural Signals}},
  author={Yang, Fengyu and Feng, Chao and Wang, Daniel and Wang, Tianye and Zeng, Ziyao and Xu, Zhiyang and Park, Hyoungseob and Ji, Pengliang and Zhao, Hanbin and Li, Yuanning and others},
  journal={arXiv preprint arXiv:2407.14020},
  year={2024}
}
\bibliographystyle{icml2026}

\newpage
\appendix
\onecolumn
\section{Caption Case}
\label{app:case}

\begin{figure*}[ht]
\begin{center}
\centerline{\includegraphics[width=0.8\linewidth]{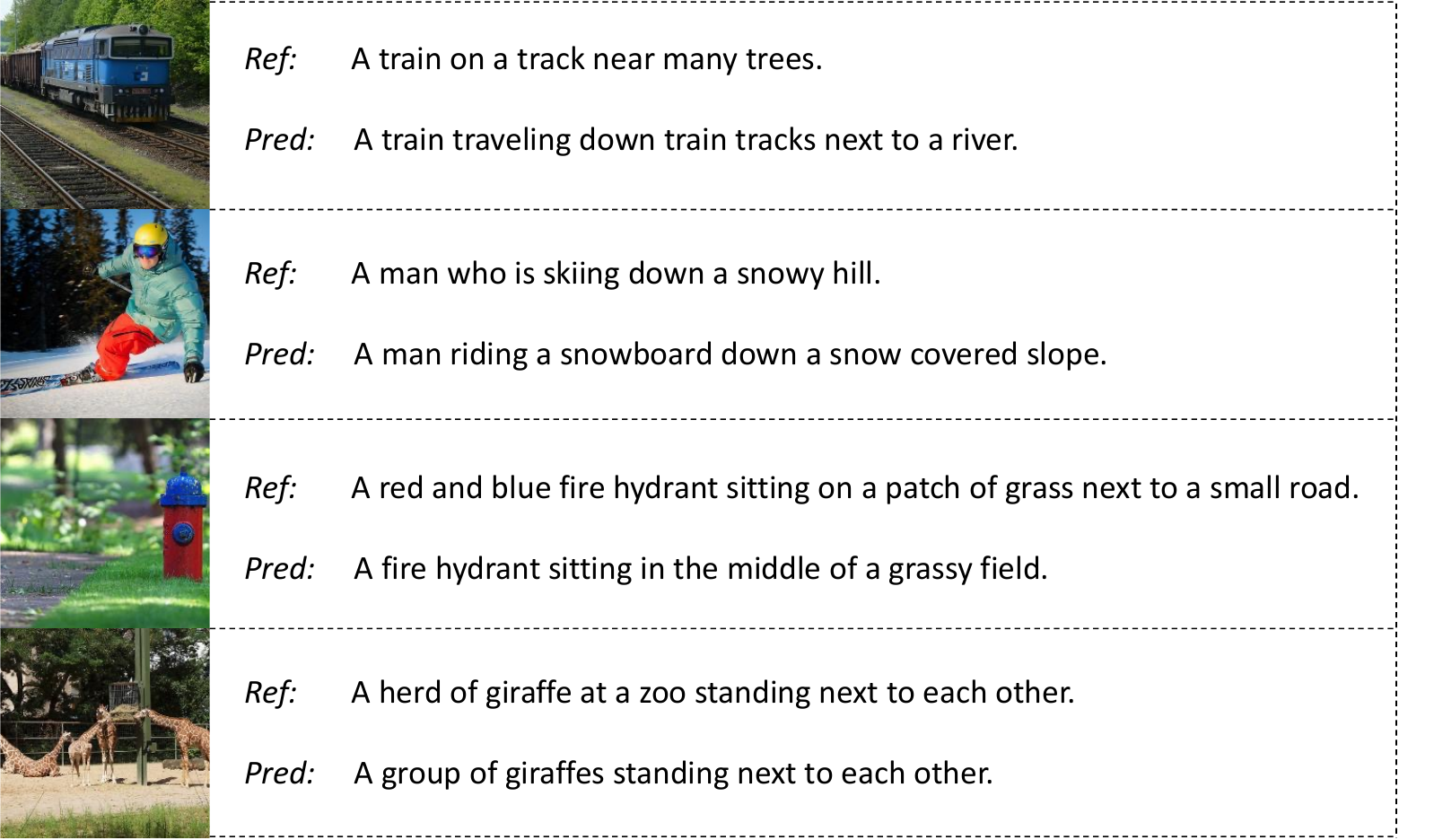}}
\caption{Examples of NOBEL decoding results on NSD. Each case displays the original visual stimulus presented to the subject alongside the reference caption and the model's predicted caption. These demonstrate that NOBEL can predict semantically accurate image descriptions directly from brain activity.}
\label{fig:nsdcase}
\end{center}
\end{figure*}

Fig~.\ref{fig:nsdcase} provides some examples of text captions generated by NOBEL from fMRI signals on the NSD dataset.

\section{Details of Datasets}
\label{app:dataset}
In this section, we provide detailed descriptions of the datasets utilized for the instruction tuning and evaluation of NOBEL:

\noindent \textbf{AD65}~\cite{ds004504:1.0.2}: AD65 consists of resting-state EEG recordings collected from 36 patients with Alzheimer’s disease and 29 healthy controls. Signals were recorded using 19 channels at 500 Hz. The dataset provides 5349 segments of 10 seconds, which are used for binary classification of Alzheimer’s disease.

\noindent \textbf{MDD}~\cite{MDD2016}: This dataset includes EEG recordings from 35 patients diagnosed with major depressive disorder and 30 healthy controls, collected using 20 channels at 256 Hz. Data were recorded under three conditions: eyes open, eyes closed, and task engagement. In total, 7302 samples of 10-second duration are used to train a classifier for distinguishing MDD patients from controls.

\noindent \textbf{PD31}~\cite{ds002778:1.0.5}: The PD31 dataset contains resting-state EEG data collected from 16 healthy controls and 15 individuals with Parkinson’s disease, using 32 channels at 512 Hz. A total of 882 segments of 10 seconds are extracted to support binary classification for Parkinson’s disease diagnosis.

\noindent \textbf{TUAB}~\cite{TUAB2016}: TUAB is a large-scale EEG dataset consisting of recordings from 2328 subjects, acquired with 21 channels at a sampling rate of 256 Hz. Each recording is annotated as either normal or abnormal. For the downstream task, 408{,}853 non-overlapping 10-second segments are extracted and used to perform binary classification of EEG normality.

\noindent \textbf{TUEV}~\cite{TUAB2016}: The TUEV dataset contains EEG signals collected from 294 individuals using 21 channels at 256 Hz. The recordings are divided into 112{,}237 segments of 5 seconds, each labeled as one of six event categories: spike and sharp wave (SPSW), generalized periodic epileptiform discharges (GPED), periodic lateralized epileptiform discharges (PLED), eye movement (EYEM), artifact (ARTF), or background activity (BCKG). A multi-class classification task is conducted to identify these event types.

\noindent \textbf{FACED}~\cite{Chen2023}: FACED consists of EEG recordings from 123 subjects watching 28 emotion-inducing video clips, covering nine emotion categories. Signals are recorded using 30 channels at either 1000 Hz or 250 Hz. For evaluation, the original labels are mapped to three coarse-grained classes (negative, neutral, and positive), yielding 10{,}332 samples of 10 seconds for emotion classification.

\noindent \textbf{WBCIC\_SHU}~\cite{Yang2025}: The WBCIC\_SHU dataset comprises high-density EEG recordings (58 channels, 1000 Hz) from 51 participants performing motor imagery tasks involving left- and right-hand grasping. A total of 30{,}591 segments, each lasting 4 seconds, are extracted for a binary classification task to recognize the imagined movement type.

\noindent \textbf{PhysioNet-MI}~\cite{Schalk2004}: PhysioNet-MI provides EEG data recorded from 109 subjects using 64 channels at a sampling rate of 160 Hz during motor imagery and motor execution experiments. The dataset includes four distinct task conditions. A total of 9837 4-second segments are employed to classify the corresponding motor tasks.

\noindent \textbf{ASD74}~\cite{ds005234:2.1.7}: ASD74 is a MEG dataset comprising recordings from 35 children diagnosed with autism spectrum disorder and 39 typically developing children. Data were acquired with 306 channels at 1000 Hz while participants watched videos accompanied by auditory stimuli. In total, 12{,}320 10-second segments are used to classify ASD versus typical development.

\noindent \textbf{MEG-MMI}~\cite{ds003568:1.0.4}: The MEG-MMI dataset includes MEG recordings acquired from 29 adolescents with major depression and 22 healthy controls during mood induction experiments. Data were recorded using 269 magnetometer channels at 1200 Hz. A total of 1770 segments, each 30 seconds long, are used to classify the presence of major depression.

\noindent \textbf{SomatoMotor}~\cite{ds006035:1.0.0}: The SomatoMotor dataset includes simultaneous EEG (74 channels) and MEG (306 channels) recordings from five subjects, sampled at 1004 Hz. During the experiment, participants received electrical stimulation to the right median nerve and were instructed to lift the left index finger as quickly as possible. A total of 1208 2-second samples are used to distinguish stimulus-induced finger movements from spontaneous ones.

\noindent \textbf{ABIDE-I}~\cite{craddock2013neuro}: The ABIDE-I dataset includes preprocessed resting state fMRI recordings from 505 individuals suffering from autism spectrum disorders (ASD) and 530 typical controls. A total of 25181 16-second samples were used for gender classification and classification to predict the presence of ASD.

\noindent \textbf{ADHD200}~\cite{bellec2017neuro}: The ADHD200 dataset includes preprocessed resting state fMRI recordings from 362 children and adolescents diagnosed with ADHD and 585 typically developing controls. A total of 23321 20-second samples were used for classification to predict the presence of ADHD.

\noindent \textbf{Neural Processing}~\cite{lepping2016neural}: The Neural Processing Dataset includes functional MRI (fMRI) recordings from 19 individuals with major depressive disorder (MDD) and 20 never-depressed (ND) control participants, who listened to standardized musical and nonmusical stimuli during scanning. A total of 3832 20-second samples were used for gender classification and classification to predict the presence of MDD.

\noindent \textbf{Working Memory}~\cite{repovvs2012working}: The Working Memory dataset includes task-based fMRI data during three working memory loads (0-back, 1-back, 2-back) of an N-back task from 23 individuals suffering from schizophrenia (SCZ) and 76 typical controls. A total of 6336 20-second samples were used for gender classification and classification to predict the presence of SCZ.

\noindent \textbf{XP1-2}~\cite{lioi2020simultaneous}: The XP1 and XP2 dataset include fMRI from right-handed, neurofeedback (NF)-naive healthy individuals  during motor imagery NF tasks. A total of 2237 20-second fMRI samples were used for gender classification.

\noindent \textbf{HCP}~\cite{hcp}: The HCP dataset includes imaging data from approximately 1,200 healthy young adults. All functional data used in this work are preprocessed and aligned to the MNI space. The resting-state fMRI data were acquired with a repetition time (TR) of 0.8 s. During scanning, participants were instructed to remain at rest. Each subject completed one or two resting-state sessions. To construct supervised learning targets, we adopt the Yeo’s 7-network cortical parcellation. For each sample, we introduce a controlled bias to regions within a selected functional network and treat the corresponding network index together with the bias magnitude as labels.

\noindent \textbf{LEMON}~\cite{lemon}: The LEMON dataset includes EEG(62-channels,  2.5kHz) and preprocessed resting state fMRI recordings from 227 subjects. A total of 11173 20-second EEG samples and 7139 20-second fMRI samples were used for gender classification.

\noindent \textbf{Nat-View}~\cite{natview}: The Nat-View dataset includes simultaneous EEG(61 channels, 5kHz) and preprocessed fMRI recordings from 22 subjects. Recordings cover tasks including 12Hz flickering checkerboard, resting state, Inscapes animation, naturalistic videos, and PEER calibration. A total of 7705 20-second EEG and fMRI samples were used for gender classification.

\noindent \textbf{SMN4Lang}~\cite{smn4lang}: The SMN4Lang dataset includes MEG(306-channels,  1000Hz) and  fMRI recordings from the same 12 subjects while the subjects listened to 6 hours of naturalistic stories. A total of 12684 20-second EEG and fMRI samples were used for gender classification.

\noindent \textbf{NSD}~\cite{allen2022massive}: The NSD dataset contains 7T fMRI recordings from eight subjects viewing natural images. The functional data have an isotropic spatial resolution of 1.8 mm and a repetition time of 1.6 s. Each subject contributed on average approximately 35.5 hours of scanning, corresponding to 26,625 stimulus trials. Event-level neural responses are estimated using a general linear model, from which beta weights are derived. In this study, we follow the dataset split in Mindeye2.

\noindent \textbf{HAD}~\cite{zhou2023large}: The HAD dataset consists of 3T fMRI recordings from 30 subjects. Functional images were acquired with an isotropic spatial resolution of 2 mm and a repetition time (TR) of 2 s. During the experiment, participants viewed short action videos spanning 180 action categories, with four distinct 2-second clips presented for each category. All subjects' functional data were spatially normalized to the MNI space. Event-related neural responses were estimated using a general linear model. The resulting representations were used for cross-subject action category classification.

All raw fMRI data were preprocessed using fMRIPrep \cite{esteban2019fmriprep}. For all datasets, the initial steps included normalizing the data to the Montreal Neurological Institute (MNI) space and standardizing each 3D volume to a 96×96×96 size with a 2×2×2 mm spatial resolution. For fMRI sequences with mismatched original spatial resolution or TR, we first resampled the data to the target resolution, then cropped or padded the volumes to unify all of them to 96×96×96.

For EMEG data, we used minimal standardized preprocessing to maximize data usability. First, a 0.1–96 Hz bandpass filter and 50/60 Hz notch filters were applied to remove slow signal drifts and power line noise; for MEG recordings with HPI coils, extra filtering was added for their corresponding HPI frequency bands. All signals were then resampled to a 250 Hz sampling rate. Bad channels were identified with a power spectral density-based detection algorithm and then interpolated for repair. To address inconsistent reference values and large amplitude fluctuations, we subtracted the reference value from all sensors of the same type for each sampling point, and normalized each channel to a zero mean and unit variance at the sampling level. Sensor coordinates and orientations were obtained from either the positions provided with the dataset or the device's standard montage.






\end{document}